\documentclass{svproc}
\usepackage[utf8]{inputenc}
\usepackage[T1]{fontenc}
\usepackage{lmodern}
\usepackage{ulem}
\usepackage{cancel}

\usepackage{amsmath,bm}

\usepackage[bottom]{footmisc}
\usepackage{cite}

\usepackage{type1cm} 

\usepackage{url}

\usepackage{xcolor}
\usepackage{microtype}

\newcommand{\fitmat}[1]{\mathbf{#1}}
\newcommand{\fitvec}[1]{\mathbf{#1}}

\definecolor{changecolor}{rgb}{0.0, 0.0, 0.0}

\begin{document}
\mainmatter

\title{A Port-Hamiltonian System Perspective on Electromagneto-Quasistatic Field Formulations of Darwin-Type}
\titlerunning{A PHS Perspective on EMQS Field Formulations of Darwin-Type} 
\author{Markus Clemens\inst{1}
\and
Marvin-Lucas Henkel\inst{1}
\and
Fotios Kasolis\inst{1}
\and
Michael Günther\inst{2}
}
\institute{Chair of Electromagnetic Theory,\\ University of Wuppertal, 42119 Wuppertal, Germany,\\
\email{\{clemens, mhenkel, kasolis\}@uni-wuppertal.de},\\ \texttt{https://tet.uni-wuppertal.de}
\and Chair of Applied and Computational Mathematics (ACM),\\ University of Wuppertal, 42119 Wuppertal, Germany,\\
\email{guenther@uni-wuppertal.de}\\
\texttt{https://acm.uni-wuppertal.de}
}

\maketitle

\begin{abstract}
  Electromagneto-quasistatic (EMQS) field formulations are often dubbed as Darwin-type field formulations which  approximate the Maxwell equations by neglecting radiation effects while modelling resistive, capacitive, and inductive effects. A common feature of EMQS field models is the Darwin-Amp\'ere equation formulated with the magnetic vector potential and the electric scalar potential. EMQS field formulations yield different approximations to the Maxwell equations by choice of additional gauge equations.
  These EMQS formulations are analyzed within the port-Hamiltonian system (PHS) framework. It is shown via the PHS compatibility equation that formulations based on the  combination of the Darwin-Amp\'ere equation and the full Maxwell continuity equation yield port-Hamiltonian systems implying numerical stability and specific EMQS energy conservation.
\end{abstract}

\section{Introduction}

\label{clemens_sec:introduction}
Electromagnetic field problems, where objects are small in comparison to the shortest wave length of the problem, belong to the regime of quasistatic field models, where electromagnetic wave propagation effects are neglected. 
In addition to the established magneto- and electro-quasistatic field models,
electro\-magneto-quasistatic field models, or so called Darwin-type field models w.r.t. to the original work in \cite{Darwin1920:01s}, have become of interest as quasistatic approximations of the full set of Maxwell equations.

These field models are of relevance for applications such as e.g. high-frequency coils or middle frequency transformers (MFTs), where resistive, capacitive and inductive field effects are considered simultaneously. 
The EMQS models of Dar\-win-type commonly feature the Darwin-Amp\`{e}re equation formulated in terms of the magnetic vector potential $\bm{A}$ and the electric scalar potential $\varphi$ as
\begin{equation}
\label{clemens_Darwin-Ampere-Time}
 \mathrm{curl}(\nu\mathrm{curl} \bm{A}) + \kappa \frac{\partial}{\partial t} \bm{A} + \kappa\mathrm{grad}\varphi + \varepsilon \mathrm{grad} \frac{\partial}{\partial t}\varphi = \bm{j}_{\mathrm{s}},
\end{equation}
\textcolor{changecolor}{where $\kappa,\varepsilon,\nu$ denote the coefficients for the electric conductivity, the permittivity and the reluctivity, respectively.}

Here, the expression $\varepsilon\frac{\partial^2}{\partial t^2}\bm{A}$ is omitted in comparison to the full Maxwell-Amp\`{e}re equation.
The Darwin-continuity equation 
\begin{equation}
\label{clemens_Darwin-Cont-Time}
 \mathrm{div}\left( \kappa\frac{\partial}{\partial t} \bm{A} + \kappa  \mathrm{grad} \varphi 
+ \varepsilon\mathrm{grad} \frac{\partial}{\partial t}\varphi\right) = \mathrm{div} \bm{j}_{\mathrm{s}},
\end{equation}
is implicitly included in \eqref{clemens_Darwin-Ampere-Time} and differs from the full Maxwell continuity equation by omission of $\mathrm{div} (\varepsilon\frac{\partial^2}{\partial t^2}\bm{A}).$

Various EMQS formulations have been derived using different continuity and gauge equations next to \eqref{clemens_Darwin-Ampere-Time}. EMQS formulations directly based on~\eqref{clemens_Darwin-Ampere-Time} and \eqref{clemens_Darwin-Cont-Time}
use added Coulomb-type gauge terms \cite{ClGu_KochSchneiderWeiland2012}. Alternatively, in \cite{HenkelKasolisClemens2023:01s} a grad-div-extension of \eqref{clemens_Darwin-Ampere-Time} was used to regularize the combination of 
\eqref{clemens_Darwin-Ampere-Time} and \eqref{clemens_Darwin-Cont-Time}.
 
Other EQMS formulations combine \eqref{clemens_Darwin-Ampere-Time} with the Maxwell continuity equation \cite{ClGu_KaimoriMifuneKameari2022, HenkelKasolisClemensGuenterSchoeps2022:01s} which yields symmetric formulations. 
Combined with the finite integration technique (FIT) \cite{Weiland1996:01s} or related mimetic discretization schemes this formulation implicitly enforces a Coulomb-type gauge $\mathrm{div}(\varepsilon\frac{\partial}{\partial t}\bm{A})=0.$ Related EMQS formulations as in \cite{ZhaoTang2019:01s} additionally and explicitly enforce this Coulomb-type gauge within Lagrange multiplier formulations. 
A number of EMQS formulations use domain dependent gauge equations and split scalar potentials to enforce a low-frequency stabilization \cite{ClGU_BadicsPavoBiliczGyimothy2023,ClGu_KaimoriMifuneKameari2022}. 

Port-Hamiltonian systems (PHS) models \cite{JacobZwart:2012} are used to maintain physical properties such as energy conservation and/or dissipative
inequalities. 
\textcolor{changecolor}{
While this contribution focuses on PHS formulations for EMQS field models, there is also interest in PHS in other areas of electromagnetic
field research, such as PHS for Maxwell equations, as outlined in references \cite{ClGu_Farle2013:01s} and \cite{HAINE2022424}, and PHS for magneto-quasistatics, as presented in references \cite{ClGu_Maciejewski2018} and \cite{ReisStykel2023}.
}

We consider port-Hamiltonian partial differential-algebraic equation (pH-PDAE) system
of the form \textcolor{changecolor}{\cite{Bartel_etal2024}}
\begin{equation}
\label{phs-pdae}
{\mathbf E}\, \frac{d}{dt} \mathbf{x} = ({\mathbf J}-{\mathbf R})\, {\mathbf z}(\mathbf{x}) + {\mathbf B} \, 
{\mathbf u}(t),\quad \mathbf{x}(0) = \mathbf{x}_0, \quad  {\mathbf y}(t) = {\mathbf B}^{\star} \mathbf{z}(\mathbf{x}).
\end{equation}
{\color{changecolor} The differential operator {$\mathbf E$ } is assumed to be { formally} self-adjoint and $\mathbf R$  { is} assumed to be formally self-adjoint and positive-semi-definite { with respect to the $L^2$-inner product $\langle \cdot, \cdot \rangle$, i.e.,
\begin{align*}
    \langle E x, y \rangle & = \langle x, E^\star y \rangle, \quad
    \langle R x, y \rangle  = \langle x, R^\star y \rangle, \quad
    \langle R x, x \rangle \ge 0.
\end{align*}
} 
In addition, we assume that the differential operator ${\mathbf J}$ is formally skew-adjoint.}

The function $\mathbf{z}$ depends on the inner state variable $\mathbf{x}$ and the input is given by $\mathbf{u}(t)$.
If the compatibility relation 
\begin{eqnarray}\label{clemens_pH-DAE_consistency_relation}
    {\mathbf E}^{\star} {\mathbf z} = \mathrm{grad} \, H 
\end{eqnarray}
holds for the Hamiltonian $H=H(\mathbf{x})$,
which represents an energy  expression,
 the dissipativity inequality $\frac{d}{dt}H(\mathbf{x})   \le {\mathbf y}^{\top} {\mathbf u}$ holds:
\begin{align}
    \tfrac{d}{dt} H(\mathbf{x}) & = 
    \color{changecolor} \langle \mathrm{grad} \, H, \dot{\mathbf{x}} \rangle  =
    \langle \mathbf{E}^\star \mathbf{z},  \dot{\mathbf{x}} \rangle  
    =  \underbrace{\langle \mathbf{z}, \mathbf{J} \mathbf{z} \rangle}_{=0} - 
    \underbrace{\langle \mathbf{z}, \mathbf{R} \mathbf{z}\rangle}_{\ge 0} + 
    \langle \underbrace{\mathbf{B}^\star \mathbf{z}}_{\fitvec{y}}, \mathbf{u} \rangle  
    \le \langle \mathbf{y},  \mathbf{u} \rangle,
\end{align}
with $\mathrm{grad} \, H$  given as the variational derivative of the Hamiltonian~\cite[Sect. 2.3]{ClGu_Maciejewski2018}. 
The full set of Maxwell's equations formulated with electrodynamic potentials 
already has a port-Hamiltonian structure~\eqref{phs-pdae} with
$\mathbf{z}=\mathbf{x}:=(\tfrac{\partial}{\partial t}\bm{A},  \varphi, \bm{H} \! )^\top$,
\begin{equation}
\label{clemens_FullMaxwell-pHs}
\begin{aligned}         
\fitmat{E}_{\mathrm{Maxwell}}=
\left[
   \begin{array}{ccc}
   \varepsilon                 &          \varepsilon \mathrm{grad} & 0\\
   -\mathrm{div}  \varepsilon    &  -\mathrm{div}  \varepsilon \mathrm{grad} & 0\\
   0 & 0  & \mu
   \end{array}
   \right], &\quad &
   \fitmat{J}= \left[ 
   \begin{array}{ccc}
   0             & 0  & -\mathrm{curl} \\
   0             & 0  &              0\\
    \mathrm{curl} & 0  & 0
   \end{array}
   \right], \\ 
   \fitmat{R} =
\left[ 
   \begin{array}{ccc}
           \kappa &          \kappa \mathrm{grad}  & 0\\
   -\mathrm{div} \kappa &  -\mathrm{div}  \kappa \mathrm{grad} & 0\\
   0    & 0 &       0
   \end{array}
   \right], & \quad &
   \fitmat{B}=
\left[
   \begin{array}{c}
   \! 1 \! \! \\
   \! -\mathrm{div}  \! \! \\
   \! 0          \! \! 
   \end{array}
   \right],
\end{aligned}
\end{equation}
\textcolor{changecolor}{with suitably chosen boundary condtions, and where $\mu=\nu^{-1}$ denotes the coefficient for the magnetic permeability. $\bm{A}, \varphi, \bm{H}$ denote} the magnetic vector potential, the electric scalar potential and the magnetic field intensity, respectively. The input $\fitvec{u}(t) = \bm{j}_{\mathrm{s}}(t)$ is the source current densities and the output $\fitvec{y}(t)= \tfrac{\partial}{\partial t} \bm{A} {\color{changecolor} + \mathrm{grad}  \varphi} (=- \bm{E})$ is the negative electric field intensity.
\textcolor{changecolor}{
The initial value $\mathbf{x}_0:=(\tfrac{\partial}{\partial t}\bm{A}_0,  \varphi_0, \bm{H}_0 \! )^\top$ must satisfy the constraint $\mu \mathbf{H}_0 = \mathrm{curl}\mathbf{A}_0$. Furthermore, if time discretization is employed, an analogous constraint must hold in a time discretized sense \cite{ClGu_ibClemensWeiland2001:01s}.
}

The Hamiltonian  of (\ref{clemens_FullMaxwell-pHs})
\begin{equation}\label{eqn_fullMaxwell-Hamiltonian}
    H=  \int_\Omega \left\{\color{black} \frac{1}{2}
     \left(\varepsilon\|\mathrm{grad}\varphi\|^2
    +  \mu \|\bm{H}\|^2 
    + \varepsilon \|\frac{\partial}{\partial t}\bm{A}\|^2 \right)
    +  \varepsilon \langle \frac{\partial}{\partial t}\bm{A},  \mathrm{grad} \varphi \rangle  \right\} d\Omega
\end{equation}
is the electromagnetic energy  with the Euclidian inner product $\langle \cdot , \cdot \rangle $.
{\color{changecolor}
We get 
\begin{align}
    \frac{d}{dt} H & = \langle \varepsilon (\mathrm{grad}  \varphi + \bm{\dot{A}}),  \mathrm{grad} \dot \varphi \rangle +
   \langle \mu \bm{H},  \bm{\dot H} \rangle +  \langle \varepsilon  \bm{\ddot{A}},  \bm{\dot{A}}
  + \mathrm{grad} \varphi \rangle  \\
  & = \langle \mathrm{grad} H, \bm{\dot{x}} \rangle
\end{align}
with 
\begin{align}
\mathrm{grad} H & =
\begin{bmatrix}
    \varepsilon (\mathrm{grad}  \varphi + \bm{\dot{A}}) \\
    -\mathrm{div} \varepsilon (\mathrm{grad}  \varphi + \bm{\dot{A}}) \\
    \mu \bm{H} 
\end{bmatrix}.
\end{align}
One easily verfies the compatability condition \eqref{clemens_pH-DAE_consistency_relation}, as we get
$
    \fitmat{E}_{\mathrm{Maxwell} }^{\star} {\mathbf z} =
    \fitmat{E}_{\mathrm{Maxwell} }{\mathbf z} = \mathrm{grad} H.
$
}
\textcolor{changecolor}{
The resulting PHS inequality 
$\frac{d}{dt}H\le \color{changecolor} \langle \fitvec{\bm y}, {\fitvec{\bm u}} \rangle =  -\langle \bm{E}, \fitvec{j}_{\mathrm{s}} \rangle $ 
corresponds to a variant of Poynting's theorem formulated without boundary radiation losses $\mathrm{div} \mathbf {S} = \mathrm{div} (\bm{E} \times \bm{H})$.
}

\section{EMQS PHS Formulations} \label{clemens_sec:phDAE-EMQS} 
Combining \eqref{clemens_Darwin-Ampere-Time} and \eqref{clemens_Darwin-Cont-Time} yields
\begin{equation}         
\label{clemens_Darwin-pHs}
\begin{split}
   \underbrace{
   \left[
   \begin{array}{ccc}
   0                 &            \varepsilon \mathrm{grad}& 0\\
   0           & \color{changecolor} -\mathrm{div}  \varepsilon \mathrm{grad}& 0\\
   0 & 0  & \mu
   \end{array}
   \right]
   }_{  {\mathbf E}_{\mathrm{EMQS1}} } 
   \frac{\partial}{\partial t}
%
  \left[
   \begin{array}{c}
    \! \frac{\partial}{\partial t}\bm{A}       \!  \\
    \! \varphi \!  \\ 
    \! \bm{H} \! 
   \end{array}
   \right]
%
   &=
%
   \left[
   \left[ 
   \begin{array}{ccc}
   0             & 0  & -\mathrm{curl}\\
   0             & 0  &              0\\
   \mathrm{curl} & 0  & 0
   \end{array}
   \right]
   \right.
%
   \\
   & \phantom{abcd}  -
   \left.
   \left[ 
   \begin{array}{ccc}
           \kappa &         \kappa \mathrm{grad}  & 0\\
   \color{changecolor} -\mathrm{div}\kappa & \color{changecolor} - \mathrm{div} \kappa \mathrm{grad}  & 0\\
   0    & 0 &       0
   \end{array}
   \right]
   \right]
     \left[
   \begin{array}{c}
    \! \frac{\partial}{\partial t}\bm{A}       \!  \\
    \! \varphi \!  \\ 
    \! \bm{H} \! 
   \end{array}
   \right] 
   \!\!
   +
   \!\!
  \left[
   \begin{array}{c}
   \! 1 \! \! \\
   \! \color{changecolor} -\mathrm{div} \! \! \\
   \! 0          \! \! 
   \end{array}
   \right] \bm{j}_{\mathrm{s}},
   \end{split}
\end{equation}
with  input $\mathbf{u}(t)=\bm{j}_{\mathrm{s}}$ and output $\mathbf{y}(t)=
\left( \frac{\partial}{\partial t}\bm{A}  +\mathrm{grad} \varphi\right) = -\bm{E}.$
Equation (\ref{clemens_Darwin-pHs}) refers to a formulation without additional gauge equation yet, i.e., a Coulomb-type gauge condition  
\begin{eqnarray}  \label{clemens_Coulomb-condition}
\mathrm{div} \left(\varepsilon \frac{\partial}{\partial t}\bm{A}\right) =0
\end{eqnarray}
is not yet enforced.

{ In this case, $\fitmat{E}_{\mathrm{EMQS1}}$ is not self-adjoint. In addition, the  compatibility condition \eqref{clemens_pH-DAE_consistency_relation} can not hold as the first row of the conjugate operator matrix $\fitmat{E}_{\mathrm{EMQS1}}^{\star}$ with only zero entries would require that a Hamiltonian  $H$ must not contain an expression $\frac{\partial}{\partial t}\bm{A}$, whereas the second row requires $\mathrm{div} (\varepsilon \frac{\partial}{\partial t}\bm{A})$ to occur in the Hamiltonian  $H$.
This argument also holds if a Coulomb-type gauge condition  
$\mathrm{div} \left(\hat{\kappa} \frac{\partial}{\partial t}\bm{A}\right) =0
$ is added to (\ref{clemens_Darwin-Cont-Time}) as in \cite{ClGu_KochSchneiderWeiland2012}
with an artificial electrical conductivity $\hat{\kappa}$ defined either in the physically non-conductive regions with $\kappa=0$ or throughout the computational domain, alternatively.
 
A symmetrization of the operator matrix $\mathbf{E}$ can be achieved by replacing the Darwin continuity equation \eqref{clemens_Darwin-Cont-Time} by the Maxwell continuity equation with  
\begin{equation}         \label{clemens_Symmetrized_Darwin-pHs}
\begin{split}
   \underbrace{
   \left[
   \begin{array}{ccc}
   0                         &              \varepsilon \mathrm{grad}& 0\\
   \color{changecolor} -\mathrm{div}  \varepsilon & \color{changecolor} -\mathrm{div} \varepsilon \mathrm{grad}& 0\\
   0                         & 0                                     & \mu
   \end{array}
   \right]
   }_{  {\mathbf E}_{\mathrm{EMQS2}} } 
   \frac{\partial}{\partial t}
%
  \left[
   \begin{array}{c}
    \! \frac{\partial}{\partial t}\bm{A}       \!  \\
    \! \varphi \!  \\ 
    \! \bm{H} \! 
   \end{array}
   \right]
   &=
%
   \left[
   \left[ 
   \begin{array}{ccc}
   0             & 0  & -\mathrm{curl}\\
   0             & 0  &              0\\
   \mathrm{curl} & 0  & 0
   \end{array}
   \right]
   \right.
%
   \\
   & - 
   \left.
   \left[ 
   \begin{array}{ccc}
           \kappa &         \kappa \mathrm{grad}  & 0\\
   \color{changecolor} -\mathrm{div}\kappa & \color{changecolor} -\mathrm{div} \kappa \mathrm{grad}  & 0\\
   0    & 0 &       0
   \end{array}
   \right]
   \right]
     \left[
   \begin{array}{c}
    \! \frac{\partial}{\partial t}\bm{A}       \!  \\
    \! \varphi \!  \\ 
    \! \bm{H} \! 
   \end{array}
   \right] 
   \!\!
   +
   \!\!
  \left[
   \begin{array}{c}
   \! 1 \! \! \\
   \! \color{changecolor} -\mathrm{div} \! \! \\
   \! 0          \! \! 
   \end{array}
   \right] \bm{j}_{\mathrm{s}},
\end{split}
\end{equation}
with  input $\mathbf{u}(t)=\mathbf{j}_{\mathrm{s}}$ and output $\mathbf{y}(t)=
( \frac{\partial}{\partial t}\bm{A}  +\mathrm{grad} \varphi) = -\bm{E}.$ 
This approach implicitly enforces the condition $\mathrm{div}(\varepsilon\frac{\partial^2}{\partial t^2}\bm{A})=0,$ which is the difference between the Darwin continuity equation \eqref{clemens_Darwin-Cont-Time} implicitly given via the Darwin-Amp\`{e}re equation \eqref{clemens_Darwin-Ampere-Time} and the full Maxwell continuity equation. Discrete versions of this symmetrized EMQS formulation are introduced in \cite{HenkelKasolisClemensGuenterSchoeps2022:01s, ClGu_KaimoriMifuneKameari2022}.  

The symmetrized EMQS formulation~\eqref{clemens_Symmetrized_Darwin-pHs}, however, yields a PHS formulation as the prerequisite compatibility relation \eqref{clemens_pH-DAE_consistency_relation} 
$
        {\mathbf E}_{\mathrm{EMQS2}}^{\star} {\mathbf z} = \mathrm{grad} \,H_{\mathrm{EMQS}}
$
{\color{changecolor}
with
\begin{align*}
\mathrm{grad} H_{\mathrm{EMQS}} & =
\begin{bmatrix}
    \varepsilon \mathrm{grad}  \varphi  \\
     -\mathrm{div} \varepsilon \frac{\partial}{\partial t}\bm{A}  \color{black} -\mathrm{div} \varepsilon \mathrm{grad}  \varphi  \\
    \mu \bm{H} 
\end{bmatrix}
\end{align*}
}
holds for the electromagneto-quasistatic approximation of the full Maxwell Ha\-miltonian  \eqref{eqn_fullMaxwell-Hamiltonian} with
\begin{equation}\label{eqn_EMQS-Hamiltonian}
    H_{\mathrm{EMQS}}=  \int_\Omega \left\{ \color{black} \frac{1}{2}
    \left(\varepsilon\|\mathrm{grad}\varphi\|^2
    +  \mu \|\bm{H}\|^2 \right)
    +  \varepsilon \frac{\partial}{\partial t}\bm{A} \cdot \mathrm{grad}\varphi \right\},
\end{equation}
corresponding to the energy  of the EMQS field approximation.

A non-symmetric gauging, i.e., adding a non-physical Coulomb-type gauge expression $\mathrm{div}\left( \hat{\varepsilon} \frac{\partial^2}{\partial t^2} \fitvec{a}\right)=0$ with a non-physical permittivity value $\hat{\varepsilon}$ to the Darwin continuity equation \eqref{clemens_Darwin-Cont-Time} (similar to the introducing a non-physical conductivity $\kappa$ in the non-conductive regions in \cite{ClGu_KochSchneiderWeiland2012}) again results in a { non-self-adjoint} matrix $\fitmat{E}_{\mathrm{EMQS 3}}$ which poses contradicting demands on a Hamiltonian \textcolor{changecolor}expression if\textcolor{changecolor}{, given $\mathbf{z}=\mathbf{x}$,} \eqref{clemens_pH-DAE_consistency_relation} were to hold.

\section{Stabilized EMQS Formulations with Auxiliary Variables}
Symmetrized EMQS field formulations that in addition explicitly enforce a Cou\-lomb-type gauge condition 
$\mathrm{div} \left( \varepsilon \frac{\partial^2}{\partial t^2}\bm{A}\right)=0,$
via Lagrange multipliers as e.g. in \cite{ZhaoTang2019:01s} are reformulated as
\begin{equation}         \label{clemens_Darwin-EMQS+LagrangeGauging}
\begin{split}
   \underbrace{
   \left[
   \begin{array}{cccc}
   0          &              \varepsilon  \mathrm{grad}& 0      & \varepsilon  \mathrm{grad}\\
    - \color{black} \mathrm{div} \varepsilon    &  - \color{black}  \mathrm{div} \varepsilon  \mathrm{grad}& 0      & 0 \\
   0          & 0                          & \mu  & 0\\
   - \color{black} \mathrm{div} \varepsilon          & 0                    & 0      & 0
   \end{array}
   \right]
   }_{{\mathbf E}_{\mathrm{EMQS3}}}
   \frac{\partial}{\partial t}
%
  \left[
   \begin{array}{c}
     \! \frac{\partial}{\partial t}\bm{A}       \!  \\
    \! \varphi \!  \\ 
    \! {\bm H} \! \\
    \! \lambda \! 
   \end{array}
   \right]
   \!\!=\!\!\
%
   \left[
   \left[ 
   \begin{array}{cccc}
   0  & 0 &  -\mathrm{curl} & 0\\
   0  & 0 &  0              & 0\\
   \mathrm{curl}  & 0 &  0  & 0\\
   0  & 0 &  0              & 0
   \end{array}
   \right]
   \right.
%
   \\
   -
   \left.
   \left[ 
   \begin{array}{cccc}
    \kappa & \kappa \mathrm{grad}                          & 0 & 0\\
    - \color{black}  \mathrm{div} \kappa &  - \color{black}  \mathrm{div}\kappa \mathrm{grad}  & 0 & 0\\
   0                   & 0                                 & 0 & 0\\
   0                   & 0                                 & 0 & 0
   \end{array}
   \right]
   \right]
   %
     \left[
   \begin{array}{c}
   \! \frac{\partial}{\partial t}\bm{A}       \!  \\
    \! \varphi \!  \\ 
    \! {\bm H} \! \\
    \! \lambda \! 
   \end{array}
   \right] 
   \!\!
   +
   \!\!
  \left[
   \begin{array}{c}
   \! 1 \! \\
   \!  - \color{black}  \mathrm{div} \! \\
   \! 0 \!\\
   \! 0 \!
   \end{array}
   \right] 
   \bm{j}_{\mathrm{s}},
\end{split}
\end{equation}
with Lagrange multiplier $\lambda$, input $\mathrm{u}(t)=\bm{j}_{\mathrm{s}}$ and $\mathrm{y}(t)=
\left( \frac{\partial}{\partial t}\bm{A}  +\mathrm{grad} \varphi\right) = -\bm{E}.$ 
System \eqref{clemens_Darwin-EMQS+LagrangeGauging} is a PHS, where the compatibility relation 
\eqref{clemens_pH-DAE_consistency_relation} holds for the Hamiltonian  \eqref{eqn_EMQS-Hamiltonian}.
The Lagrange multiplier $\lambda$ corresponds to a second scalar electric potential and as solutions to \eqref{clemens_Darwin-EMQS+LagrangeGauging} exists, where $\mathrm{div} \left( \varepsilon \frac{\partial^2}{\partial t^2}\bm{A}\right)=0,$ holds and $\lambda=0,$ it is possible to modify  \eqref{clemens_Darwin-EMQS+LagrangeGauging} to a formulation that corresponds to a split potential formulation for the electric field $\bm{E}=-\frac{\partial}{\partial t}\bm{A}  -\mathrm{grad} (\varphi + \lambda)$ as introduced in \cite{HiptmairKramerOstrowski2008:01} via  
\begin{equation}         \label{clemens_Darwin-EMQS+SplitPotential}
\begin{split}
   \underbrace{
   \left[
   \begin{array}{cccc}
   0   &  \varepsilon  \mathrm{grad}& 0   & \varepsilon  \mathrm{grad}\\
   \color{changecolor} - \mathrm{div} \varepsilon & \color{changecolor} - \mathrm{div} \varepsilon  \mathrm{grad}& 0      & \color{changecolor} - \mathrm{div} \varepsilon  \mathrm{grad} \\
   0          & 0                    & \mu  & 0\\
   \color{changecolor} - \mathrm{div} \varepsilon          & \color{changecolor} - \mathrm{div} \varepsilon  \mathrm{grad}                     & 0      & \color{changecolor} - \mathrm{div} \varepsilon  \mathrm{grad}
   \end{array}
   \right]
   }_{{\mathbf E}_{\mathrm{EMQS4}}}
   \frac{\partial}{\partial t}
%
  \left[
   \begin{array}{c}
     \! \frac{\partial}{\partial t}\bm{A}       \!  \\
    \! \varphi \!  \\ 
    \! \bm{H} \! \\
    \! \lambda \! 
   \end{array}
   \right]
   \!\!=\!\!\
%
   \left[
   \left[ 
   \begin{array}{cccc}
   0              & 0 &  -\mathrm{curl} & 0\\
   0              & 0 &  0              & 0\\
   \mathrm{curl}  & 0 &  0              & 0\\
   0              & 0 &  0              & 0
   \end{array}
   \right]
   \right.
%
   \\
   -
   \left.
   \left[ 
   \begin{array}{cccc}
    \kappa             & \kappa \mathrm{grad}              & 0 & 0\\
   \color{changecolor} - \mathrm{div} \kappa & \color{changecolor} - \mathrm{div}\kappa \mathrm{grad}  & 0 & 0\\
   0                   & 0                                 & 0 & 0\\
   0                   & 0                                 & 0 & 0
   \end{array}
   \right]
   \right]
   %
     \left[
   \begin{array}{c}
   \! \frac{\partial}{\partial t}\bm{A}       \!  \\
    \! \varphi \!  \\ 
    \! \bm{H} \! \\
    \! \lambda \! 
   \end{array}
   \right] 
   \!\!
   +
   \!\!
  \left[
   \begin{array}{c}
   \! 1 \! \\
   \! \color{changecolor} - \mathrm{div} \! \\
   \! 0 \!\\
   \! \color{changecolor} - \mathrm{div} \!
   \end{array}
   \right] 
   \bm{j}_{\mathrm{s}},
\end{split}
\end{equation}
with Lagrange multiplier $\lambda$, input $\mathrm{u}(t)=\bm{j}_{\mathrm{s}}$ and $\mathrm{y}(t)=
\left( \frac{\partial}{\partial t}\bm{A}  +\mathrm{grad} (\varphi+\lambda)\right) = -\bm{E}$ as output.
With the Hamiltonian 
\begin{multline}\label{eqn_fullMaxwell-Hamiltonian2}
    H_{\mathrm{EMQS}}=  \int_\Omega \left\{ \color{black} \frac{1}{2}
    \left(\varepsilon\|\mathrm{grad}\varphi\|^2
    + \varepsilon\|\mathrm{grad}\lambda\|^2
    +  \mu \|\bm{H}\|^2 \right) \right. \\
    \left. +  \varepsilon \frac{\partial}{\partial t}\bm{A} \cdot \mathrm{grad}(\varphi+\lambda)
    +  \varepsilon \mathrm{grad}\varphi \cdot \mathrm{grad}\lambda  \right\} d\Omega
\end{multline}
corresponding the EMQS energy, \eqref{clemens_pH-DAE_consistency_relation} holds for the formulation \eqref{clemens_Darwin-EMQS+SplitPotential} with 
{\color{changecolor}
\begin{eqnarray*}
\mathrm{grad}\, H_{\mathrm{EMQS}}
= \fitmat{E}_{\mathrm{EMQS4} }^{\star} {\mathbf z} 
=\left[
   \begin{array}{c}
\varepsilon \mathrm{grad} (\varphi+\lambda), \,\\
\color{changecolor} - \mathrm{div} \left(\varepsilon (  \frac{\partial}{\partial t}\bm{A} + \mathrm{grad} (\varphi+\lambda))\right) \,\\
\mu\bm{H}\\
\color{changecolor} - \mathrm{div} \left(\varepsilon (  \frac{\partial}{\partial t}\bm{A} + \mathrm{grad} (\varphi+\lambda))\right)
   \end{array}
\right]
.
\end{eqnarray*}
}
Hence, the EMQS formulation \eqref{clemens_Darwin-EMQS+SplitPotential} is also a PHS. It should be noted that the gauge condition $\mathrm{div} \left( \varepsilon \frac{\partial^2}{\partial t^2}\bm{A}\right)=0$ can be enforced domain dependent in \eqref{clemens_Darwin-EMQS+SplitPotential}, e.g. only in non-conducting regions.
This establishes a correlation between the stabilized Lagrange multiplier EMQS formulation and various split electric scalar potential EMQS formulations as presented e.g. in
\cite{ClGU_BadicsPavoBiliczGyimothy2023} and \cite{KaimoriMifuneKameariWakao2023:01s}. 

Symmetrized EMQS field formulations can also be set up to explicitly enforce the Coulomb-type gauge condition \eqref{clemens_Coulomb-condition} via
$\mathrm{div} \left( \hat{\kappa} \frac{\partial}{\partial t}\bm{A}\right)= \frac{1}{\tau }\mathrm{div} \left( \varepsilon \frac{\partial}{\partial t}\bm{A}\right)=0,$ with an artificial $  
\hat{\kappa}:=\frac{1}{\tau }\varepsilon$ (for a time constant $\tau$ to get the correct units) 
as
\begin{equation}         \label{clemens_Darwin-EMQS+LagrangeGauging2}
\begin{split}
   \underbrace{
   \left[
   \begin{array}{cccc}
   0                       &  \varepsilon  \mathrm{grad}         & 0   & 0\\
   \color{changecolor} - \mathrm{div}\varepsilon & \color{changecolor} - \mathrm{div}\varepsilon\mathrm{grad}& 0   & 0\\
   0                       & 0                                   & \mu & 0\\
    0                      & 0                                   & 0   & 0
   \end{array}
   \right]
   }_{{\mathbf E}_{\mathrm{EMQS5}}}
   \frac{\partial}{\partial t}
%
  \left[
   \begin{array}{c}
     \! \frac{\partial}{\partial t}\bm{A}       \!  \\
    \! \varphi \!  \\ 
    \! \bm{H} \! \\
    \! \lambda \! 
   \end{array}
   \right]
   \!\!=\!\!\
%
   \left[
   \left[ 
   \begin{array}{cccc}
   0  & 0 &  -\mathrm{curl} & {  -\hat{\kappa}            \mathrm{grad}}\\
   0  & 0 &  0              & 0\\
   \mathrm{curl}  & 0 &  0  & 0\\
   { - \mathrm{div} \hat{\kappa}}   & 0 &  0              & 0
   \end{array}
   \right]
   \right.
%
   \\
   -
   \left.
   \left[ 
   \begin{array}{cccc}
    \kappa & \kappa \mathrm{grad}  & 0 &  0\\
   \color{changecolor} - \mathrm{div} \kappa & \color{changecolor} - \mathrm{div}\kappa \mathrm{grad}  & 0 & 0\\
   0                 & 0                      & 0 & 0\\
   0  & 0  & 0 & 0
   \end{array}
   \right]
   \right]
   %
     \left[
   \begin{array}{c}
   \! \frac{\partial}{\partial t}\bm{A}       \!  \\
    \! \varphi \!  \\ 
    \! \bm{H} \! \\
    \! \lambda \! 
   \end{array}
   \right] 
   \!\!
   +
   \!\!
  \left[
   \begin{array}{c}
   \! 1 \! \\
   \! \color{changecolor} - \mathrm{div} \! \\
   \! 0 \!\\
   \! 0 \!
   \end{array}
   \right] 
   \bm{j}_{\mathrm{s}},
\end{split}
\end{equation}
with the Lagrange multiplier $\lambda$, input $\mathrm{u}(t)=\bm{j}_{\mathrm{s}}$ and $\mathrm{y}(t)=
\left( \frac{\partial}{\partial t}\bm{A}  +\mathrm{grad} \varphi\right) = -\bm{E}.$ Note that $\hat{\kappa}$ can also be defined domain dependent, i.e., the gauge \eqref{clemens_Coulomb-condition} can be defined in the whole computational domain or just in the non-conducting domains with the original electric conductivity $\kappa=0.$
The system \eqref{clemens_Darwin-EMQS+LagrangeGauging2} is a PHS, where the compatibility relation 
\eqref{clemens_pH-DAE_consistency_relation} holds for the Hamiltonian  \eqref{eqn_EMQS-Hamiltonian}. { It maintains stability and energy conservation also for the case $\kappa = 0$, i.e., in the absence of ohmic losses.}
With the explicitly enforced Coulomb-type gauge \eqref{clemens_Coulomb-condition}, the Hamiltonian  \eqref{eqn_EMQS-Hamiltonian} of the symmetric formulation \eqref{clemens_Darwin-EMQS+LagrangeGauging} reduces to the expression
$
    H_{\mathrm{EMQSR}}= \frac{1}{2}
    \left(\varepsilon\|\mathrm{grad}\varphi\|^2
    +  \mu \|\bm{H}\|^2 \right).
%
$
This expression corresponds the energy of the EMQS field approximation, i.e., a formulation with the combined magnetic field and electro-quasistatic field energy.

\section{Conclusions} \label{clemens_section:conclusion}
Various infinite dimensional EMQS formulations have been classified within the port-Hamil\-tonian system framework, thus implying their numerical stability after a proper mimetic discretization where the PHS structure is found.
The validity of the compatibility condition could be shown for symmetrized formulations, i.e., those EMQS formulations based on the Darwin-Ampere and the full Maxwell continuity equation. However, this re-introduced the second order time derivative of the magnetic vector potential that was discarded in the Darwin-Ampere equation under the no radiation EMQS assumption.  For explicitely gauged EMQS formulations the relation between Lagrange multiplier and split scalar potential formulations is shown.
\subsubsection*{\ackname}
This work is supported in parts by the Deutsche For\-schungs\-gemein\-schaft DFG under grant no. CL143/18-1.
\bibliographystyle{spmpsci}
\bibliography{clemens_references}
\end{document}